
\input harvmac
\magnification=1200
\Title{\vbox{\baselineskip12pt\hbox{INRNE-TH/1/93}
\hbox{hep-th/9303087}\hbox{Revisedversion}}}
{\vbox{\centerline{Extended Non-Abelian Gauge Symmetries}
   \vskip2pt\centerline{in Classical WZNW Model}}}


\centerline{Raiko P. Zaikov\footnote{$^\star$}
{Supported by Bulgarian Fondation on Fundamental Research under
contract Ph-11/91-94}}
\bigskip\centerline{Institute for Nuclear Research and Nuclear Energy}
\centerline{Boul. Tzarigradsko Chaussee 72, 1784 Sofia}
\centerline{e-mail: zaikov@bgearn.bitnet}

\vskip .3in

\noindent{\bf Abstract. } Higher spin extensions of the
nonabelian gauge symmetries for the classical WZNW model are
considered. Both linear and nonlinear realizations of the
extended affine Kac-Moody algebra are obtained. A characteristic
property of the WZNW model is that it admits a higher spin
linear realization of the extended affine Kac-Moody algebra which
is equivalent to the corresponding higher spin nonlinear
realization of the same algebra.  However, in both cases the
higher spin currents do not form an invariant space with respect
to their generating transformations. This makes it imposible for
this symmetry to be gauged.

\Date{3/93} 


\newsec{Introduction}

\noindent After the Zamolodchikov's paper \ref\rZA{Zamolodchikov,
A. B, Theor.  Math. Phys., 63 (1985) 1205.} there is an increasing
interest in the higher spin extensions of the Virasoro algebra on
both pure Lie algebraical level, and on field theoretical
level.  A complete list of references can be found in the
Bouwknegt and Schoutens Review \ref\rBS{Bouknegt, P. and
Schoutens, {"\cal W}-{\it Symmetry in Conformal Field Theory}",  CERN
preprint CERN-TH.6583/92, to be published in Phys.  Reports}. We
recall, that one possible method for obtaining an extended
conformal algebra is to apply the operator algebra formalism.
Using this method we add to the energy-momentum tensor a finite
(infinite) number of higher spin operators, which form a closed (in
generally nonlinear) algebra \rZA . In the case when an infinite
set of higher spin generators is added we have a closed Lie
algebra.

Extended conformal algebra can be also obtained in a more
systematic way by applying the Drinfeld-Sokolov (D-S) reduction
procedure to the degree of freedom of a theory whose structure is
based on a Lie algebra \ref\rGD{Gel'fand, I. M and Dickey, L. A.,
{\it "A Family of Hamiltonian Structures Connected with Integrable
Non-Linear Differential Equations"}, IPM preprint, AN SSSR, Moskow
1978} \nref\rDS{Drinfeld, V. and Sokolov, V., J. Sov. Math., 30
(1984) 1975}--
\ref\rFL{Fateev, V. A. and Lukyanov, S. L., Int. J. Mod. Phys.,
A3 (1988) 507 \semi
Sov. Phys. JETP, 67 (1988) 447 \semi
Sov. Sci. Rev. A Phys., 15 (1990) 1.}.

Another aproach which can be called a hidden symmetry approach
has been applyied also. In this approach we start with some
two-dimensional field theoretical model with conformal invariant
Lagrangian and try to find extended symmetries providing
additional (to the energy-momentum tensor) conserved quantities
\ref\rBBSS{Bais, F. A., Bouwkned, P., Schoutens, K. and Surridge,
M., Nucl. Phys., B 304 (1988) 376.}\ (see also \ref\rZa{Zaikov,
R. P., Lett. Math.  Phys.,7 (1983) 363}, \ref\rZb{Zaikov, R. P.,
Sov. Elem. Part. and Nucl., 16 (1985) 1053.}). In the case of
free field models  such a construction coincides with the D-S Lie
algebra reductions. More interesting are the self-interacting
models as for instance the WZNW model. For that model the
extended quantum algebra based on the algebras $A_l, D_l$ and
$E_l$ was proposed in \rBBSS .  In this approach it is possible
to gauge the extended symmetry ($W$-symmetry) by including the
Noether coupling through higher spin gauge fields, i.e. to
construct $W$-gravity
\ref\rHu{Hull, C. M., Phys. Lett., B240 (1990) 110 \semi\ Phys.
Lett., B259 (1991) 68\semi\ Nucl. Phys., B353 (1991) 707 \semi\
Nucl. Phys., B364 (1991) 621.}, \nref\rBPR{Bergshoeff, E., Pope,
C. N., Romans, L. J., Sezgin, E., Shen, X. and Stelle, K. S.,
Phys. Lett., B243 (1990) 350.}--\ref\rNPV{Nissimov, E., Pacheva,
S. and Vaysburg, I., Phys. Lett., B284 (1992) 321.}.

We remind that the affine Kac-Moody algebra also admits nonlocal
(being also nonlinear) realizations which are obtained as
symmetry of the nonlinear sigma models \ref\rDR{Dolan, L. and
Roos, A., Phys.  Rev., D22 (1980) 2018}\nref\rD{Dolan, L., Phys.
Rev. Lett., 47 (1981) 1371\semi Phys. Reports, 109 (1984)
1.}\nref\rWu{Wu, Y.  S., Nucl. Phys., B211 (1983)
160.}\nref\rZa{Zaikov, R. P., Theor.  Math.  Phys., 48 (1981)
34.}--\ref\rZaaa{Zaikov, R. P., Theor.  Math. Phys., 53 (1982)
53.} and \rZb .

In the present article the hidden symmetry approach is applied
for the search of a higher spin extension of the affine Kac-Moody
algebra in case of the classical WZNW model. Both linear
and nonlinear realizations of the extended affine
Kac-Moody algebra are considered. It is shown that the higher
spin extended $SU(N)$ or $SL(N)$ affine algebras can not be closed
alone. To have a closed Lie algebra it is nesessary to include the
corresponding extended $U(1)$ algebras too.  A similar situation
appears also in the case of free fermionic model
\ref\rAA{Abdalla, E., Abdalla, M., C., B., Sotkov, G. and
Stanishkov, \ {\it "Of critical current algebra"} \ preprint
IFUSP/1027/93, Sao Paulo 1993; hep-th/9302002}. These higher spin
$U(1)$ extended algebras coincide with the classical $W_\infty $
algebra in linear or nonlinear realization (see \ref\rP{Pope, C.
N., Romands, L. J.  and Shen, X., Phys. Lett.  236B (1990) 173
and 242B (1990) 401. \semi Nucl. Phys., B339 (1990) 191.}).

The characteristic property of the classical WZNW model is that
the obtained here nonlinear realization of the extended affine
Kac-Moody algebra which we denote by $KW_\infty $ is equivalent
to the corresponding linear realization. The same property
possesses the classical algebra $W_\infty $ wich is a subalgebra
of the $KW_\infty $ algebra.

Another characteristic property of the WZNW model is that the
corresponding classical $KW_\infty $ and $W_\infty $ conserved
Noether currents do not form an invariant space with respect to
their generating transformations. This property makes it
imposible to gaige the $KW_\infty $ symmetry, as well as, the
$W_\infty $ symmetry for the classical WZNW model \rHu ,\rBPR .
Moreover, this noninvariance of the conserved currents spaces
shows that there exists some nonequivalence between the WZNW
model and the corresponding nonabelian free fermionic model
\ref\rW{Witten, E., Comm. Math. Phys., 92 (1984) 455.} on a
higher spin symmetry level.

The nonlinear realization of the classical $W_\infty $ algebra
considered here can be found by suitable deformation (the
deformation parameter $q$ is seted to $1$) of the classical
$w_\infty $ algebra \ref\rA{Aratyn, H., Ferreira, L. A., Gomes,
J. F. and Zimerman, A. H., Phys. Lett., B254 (1991) 372, \semi
{\it "A new deformation of W-infinity and application to the
two-loop WZNW model and conformal affine Toda model"} preprint
IFT-P.003/92}. The corresponding linear realization of the
$W_\infty $ coincides with the $DOP(S^1)$ Lie algebra
\ref\rRad{Radul, A. O., Pis'ma Zh. Eksp. Teor. Fis. 50 (1989)
341, \semi Phys. Lett., B265 (1991) 86}.

At the end the nolocal realization of the affine Kac-Moody
algebra \rD , \rWu , \rZb \  is also considered. It is shown that
only the zero modes of the nonlocal generators form a closed Lie
algebra.

\newsec{Ordinary symmetries in classical WZNW model}

\noindent We consider the classical action for the level 1 WZWN model:
\eqn\ea{S={1\over 4\pi }\int d^2xtr\{j^\mu (x)j_\mu (x)\}-\Gamma ,}
where
\eqn\eaa{\Gamma ={1\over 12\pi }\int d^3y\epsilon ^{\alpha \beta \gamma }
tr\{\tilde \jmath _\alpha (y)\tilde \jmath _\beta (y)
\tilde \jmath _\gamma (y)\}}
and
\eqn\eaaa{\eqalign
{&j_\mu (x)=g^{-1}(x)\partial _\mu g(x),\ \ (\mu =0,1)\cr
&\tilde \jmath _\alpha (y)=\tilde g^{-1}(y)\partial _\alpha \tilde
g_\alpha (y), \ \ (\alpha =0,1,2).\cr}}
Here the fields $g$ and $\tilde g$ take their values in some Lie
group $G$, while the currents $j$ and $\widetilde \jmath $ take
their values in the corresponding Lie algebra ${\cal G}$.

The action \ea \  is invariant
with respect to both left-acting and right-acting rigid gauge
transformations
\eqn\eab{g(x)\rightarrow h_Lg(x)h_R,}
where $h_L, h_R\in G$. It is evident that the currents \eaaa \
are invariant with respect to the left-acting rigid gauge
transformations. There are also right invariant currents:
\eqn\eabb{j_\mu \rightarrow \widehat \jmath =gjg^{-1}, \qquad
\tilde \jmath \rightarrow \widehat {\tilde \jmath }=
\tilde g \tilde {\jmath }\tilde g^{-1}.}
Important for our considerations is the symmetry of the
action (up to the sign of $\Gamma $) with respect to the change
$j\leftrightarrow \widehat \jmath $. We note, that we find \eabb
\ from the
Drinfeld-Sokolov linear system
$$ (\partial +j)\eta =0, \qquad
\partial \widehat \eta -\widehat \eta \widehat j)=0
$$
by incertion \ $\eta =g^{-1}, \widehat \eta =g$.

Now, we consider an arbitrary local infinitesimal gauge
transformation $g\rightarrow g+\delta g$. It is easy to see
that the currents \eaaa \  transform according to the gauge
potential law:
\eqn\ebbb{\eqalign{\delta _Rj_\mu & =\partial _\mu \Omega +\lbrack j_\mu
,\Omega \rbrack , \cr
\delta _L\hat \jmath _\mu & =
\partial _\mu \widehat {\Omega }+\lbrack \widehat {\jmath }_\mu ,
\widehat {\Omega }\rbrack , \cr
\delta _R\hat \jmath _\mu  & =g^{-1}\partial_\mu \Omega g, \cr
\delta _Lj_\mu  & =g^{-1}\partial _\mu \widehat {\Omega }g,\cr}}
with respect to the right-acting gauge transformations
\eqn\eac{\delta g(x)=g(x)\Omega (x).}
Here $\widehat \Omega =g\Omega g^{-1}=\delta gg^{-1}$ are the
generators of the left-acting gauge transformations.
Consequently, $j$ has the properties of a nonabelian gauge
potential with respect to the right-acting transformation $\Omega
$ while $\widehat \jmath $ has the same properties with respect
to the left-acting transformations $\widehat \Omega $.

The invariance condition for the action \ea \ with
respect to the right-acting gauge transformations \eac
\ is given by:
\eqn\ec{\delta_RS\sim \int d^2xtr\{j_+(x)\partial _-\Omega(x)\}=0,}
where the light-cone coordinates $x_\pm =(x_0\pm x_1)/2$ are
introduced and the zero curvature condition:
\eqn\eacc{\partial _\mu j_\nu -\partial _\nu j_\mu
+\lbrack j_\mu ,j_\nu \rbrack =0,}
which is a direct consequence of \eaaa \ is used. For the
derivation of the formula \ec \ we also use the fact that the
variation of the Wess-Zumino term in \ea \  can be
represented as a total divergence \rW .

In what follows we shall consider only the right-acting
transformations.

If we set in \eac :
\eqn\ebcc{\Omega =i\alpha (x_+){\cal C},}
where the constant matrix ${\cal C}\in {\cal G}$ and $\alpha (x_+)$ is
an arbitrary function of $x_+$ only, we find the generators of the
classical (without central term) affine Kac-Moody algebra.

We remind that the WZNW model with the action \ea \ is
equivalent to the free fermionic model
\eqn\eb{S_f=\int d^2x(\psi _+\partial _-\psi _+
+\psi _-\partial _+\psi _-),}
where the Majorana spinor field $\psi $ transform under the
adjoint representation of the $SO(N)$ group \rW . We note that
according to Witten's assertion \rW \ the action \eb \ has the
same symmetry properties as the action
\ea \ and that the nonabelian bosonization procedure
consists of
\eqn\ecb{j_+=\psi _+\psi _+, \qquad  \widehat{\jmath }_-=
\psi _-\psi _-, \qquad  g_{mn}(x)=\psi _{+,m}\psi _{-,n},}
where $j_+$ and $\widehat{\jmath}_-$ are given by \eabb . The
equations of motion in the both cases are
\eqn\eaca{\partial _-j_+=0,
 \qquad \qquad  \partial _+\widehat {\jmath }_-=0.}
where $j_\pm $ are given by \eabb \ or \ecb .

\newsec{Linear realizations of the extended affine Kac-Moody algebras}

\noindent In order  to  consider  extended  affine Kac-Moody
algebras we assume that  in formula
\eac \ $\Omega $ \ is replaced by a polynomial  of
$j, \partial j, \partial ^{2}j\dots $,
i. e.  we  have  the  following transformation law
\eqn\ef{\delta g(x)=
g(x)\epsilon \alpha ^a(x){\cal P}(j,\partial j,\partial^2j,\dots )t_a,}
where $\epsilon $ is a constant infinitesimal parameter, $t_a$ are
the matrix generators of the group \ $G$ \ and \ $\alpha ^a$ \ are
arbitrary functions of one of the light-cone variables only.  If
\ ${\cal P}$ \ contains the finite order derivatives only then \ef \
are  local gauge transformations.  The nonlocal  gauge
transformations were considered already as a hidden symmetry of
the sigma models and a nonlocal realization of the affine
Kac-Moody algebra was  found in \rZb , \rDR --\rZaaa .  Here we
shall consider both  local   and  nonlocal transformations as
well as a linear and a nonlinear realizaions of the higher spin
extended affine Kac-Moody algebra.


As  in   the case   of the extended Virasoro transformations
\ref\rZai{Zaikov, R. P., {\it "Extended symmetries of the classical
WZNW model"}, in preparation} (see also \rZaaa \ and \rZb ) we
have a class of higher spin extended affine Kac-Moody ($KW_\infty
$) transformations which form an off-shell symmetry of the action
as well as $KW_\infty $ transformations  which are  only on-shell
symmetry. We  try to find the off-shell symmetries as a special
nonlinear extension of the ordinary affine Kac-Moody
transformations, while all other nonlinear transformations, as
well as all linear extended affine Kac-Moody transformations are
only on-shell symmetry of the action \ea .  First  we  draw
atention  to  the off-shell symmetry. As it is shown in \rZai \
we  have an off-shell symmetry if ${\cal P}$ in \ef
\ depends on $j$ only (derivative terms are not included).
However,  the presence  of the  matrix generators $t_{a}$ in \ef \
is more restrictive    than the off-shell $U(1)$ gauge
transformations \rZai .  It  is easy   to check that   the
transformations
\eqn\efa{\delta ^m_\alpha g(x)=
ig(x)\sum^ m_{p=0}(j)^pt_a(j)^{m-p}\alpha _a^m(x_+)}
are off-shell symmetry of \ea \ (up to a total divergence term).
We note also that although the nonlinear transformations \efa \
are off-shell symmetry of the action \ea \ they  do  not  form a
closed Lie algebra.

To find a closed Lie algebra as a higher spin extension of
the ordinary affine Kac-Moody algebra we consider the nonabelian
extended  gauge transformations which are  only  on-shell
symmetry of the action and which form a closed Lie algebra. For
this purpose we start from the transformation:
\eqn\efaa{\delta ^1(\hat \alpha )g(x)=
i\partial g(x)\hat \alpha (x)=
-i\alpha ^a(x)gg^{-1}\partial gt_a.}
where $ \hat \alpha =\alpha ^at_a \ {\rm and} \  \alpha ^a$ is an
arbitrary function of one light-cone variable \ $x_+$ \ only. The
formula \efaa \ demonstrates the equivalence (on a spin $2$
level) between the linear and the nonlinear realizations of the
$KW_\infty $  transformations within the WZNW model. In the
present section we draw attention to the linear realizations
only. The nonlinear transformations will be considered in the
next section.

Now it is straightforward to derive the commutation relation
between two infinitesimal linear transformations \efaa :
\eqn\efab{\eqalign{
[\delta ^1(\hat \alpha ),\delta ^1(\hat \beta )]g(x)
&=i^2\partial ^2g[\hat \alpha ,\hat \beta ]+
i^2\partial g(x)(\partial \hat \alpha \hat \beta -
\partial \hat \beta \hat \alpha ) \cr
&=i^2f_{ab}^c\alpha ^a\beta ^b\partial g(x)t_c+
i^2\beta ^b\partial \alpha ^a\partial g(x)t_at_b-
i^2\alpha ^a\partial \beta ^b\partial g(x)t_bt_a \cr }}
First we consider the case when \ $t_a$ \ are the generators of the
\ $SU(N)$ \ or the \ $SL(N)$ \ group in the fundamental
representation for which we have
\eqn\efaba{t_at_b={1\over 2}f_{ab}^ct_c+{1\over 2}d_{ab}^ct_c+
\delta _{ab}I,}
where $f_{abc}={1\over 2}tr(\lbrack t_a,t_b\rbrack t_c)$, $d_{abc}=
{1\over 2}tr(\{t_a,t_b\}t_c)$, $tr(t_at_b)=2\delta _{ab}$ and
$I$ \ is the identity matrix. In this case the formula \efab
\ reduces to
\eqn\efabr{\eqalign{
[\delta ^1(\hat \alpha ),\delta ^1(\hat \beta )]g(x)
&=i\delta ^2([\hat \alpha ,\hat \beta ])g(x)+
{i\over 2}\delta ^1\bigl(f_{ab}^c\partial (\alpha ^a\beta
^b)t_c\bigr)g(x) \cr
 & +{i\over 2}\delta ^1\bigl(d_{ab}^c(\partial \alpha ^a\beta ^b-
\alpha ^a \partial \beta ^b)t_c\bigr)g(x)
+i\widetilde \delta ^1(\beta _a\partial \alpha ^a-
\alpha ^a\partial \beta _a)g,\cr }}
where
\eqn\efac{\delta ^2(\alpha )g(x)
=i\alpha ^a(x)\partial ^2g(x)t_a,}
and
\eqn\efad{\widetilde \delta ^1(k)g=ik(x)\partial g(x).}
The transformations \efac \ we call spin-$3$ nonabelian gauge
transformations becase they generate the spin-$3$  $SU(N)$ current
$J^3_a=tr\bigl(\partial g^{-1}\partial ^2gt_a\bigr)$, while the
transformations \efad \ are the well known Virasoro
transformations generating the spin-$2$ \ $U(1)$ current
(energy-momentum tensor) $V^0=tr(\partial g^{-1}\partial g)$. We
note, that the transformations \efad \ have the form of
a higher spin extension of the abelian  gauge transformations
$\widetilde \delta g=ikg$.

If \ $t_a$ \ are generators of the group $G=SO(N)$ in the adjoint
representation then the equation \efaba \ is not satisfyed. It
is replaced by
\eqn\efae{t_at_b={1\over 2}f_{ab}^ct_c+t_{ab},}
where $t_{ab}=\{t_a,t_b\}/2$ is symmetric but not traceless
second rank tensor.  Incerting \efae \ into \efab \ we obtain
\eqn\efaw{\eqalign{
[\delta ^1(\hat \alpha ),\delta ^1(\hat \beta )]g(x)
&=i\delta ^2([\hat \alpha ,\hat \beta ])g(x) \cr
& +{i\over 2}\delta ^1\bigl(f_{ab}^c\partial (\alpha ^a\beta ^b)t_c\bigr)g(x)+
+i\delta ^1_2\bigl((\partial \alpha ^a\beta ^b-
\alpha ^a \partial \beta ^b)t_{ab}\bigr)g(x), \cr }}
where we denote
\eqn\efaf{\delta^1_2(\alpha ^a\beta ^bt_{ab})g=i\alpha ^a\beta ^bgt_{ab}.}
In this case $(G=SO(N))$ in the r.h.s. of \efaw \ the internal
symmetry tensor terms appear also. Consequently, we have an
extension of the affine Kac-Moody algebra not only with respect
to the conformal spin, but also with respect to the internal
symmetry spin.

In what follows we restrict our considerations to the $SU(N) \
{\rm and } \ SL(N)$ \ cases for which internal higher spin
transformations do not appear. The case when such transformations
appear ($G=SO(N)$) will be considered elsewhere. We note, that in
a recent paper \rAA \  it was shown that the similar properties
possesses
the free fermionic model too.

Repeatedly applying the spin $2$ nonabelian transformations \efaa
\ we can find a nonabelian transformation of any spin which reads:
\eqn\efb{\delta ^m(\hat \alpha )g(x)=
ig(x)U^{m}(x)t_a\alpha _m^a(x)=
i\alpha _m^a(x)\partial ^mg(x)t_{a}, \qquad  (m=0,1,\dots)}
Here we have introduced the following quantities:
\eqn\efbn{U^m(x)=g^{-1}(x)\partial ^{m}g(x), \qquad (m=0,1,\dots ).}

The commutator of two infinitesimal nonabelian  gauge
transformations of arbitrary spin is given by
\eqn\efd{\eqalign{[\delta ^m(\hat \alpha ),\delta ^n(\hat \beta )]g(x)=
i\sum ^{max(m,n)}_{r=0}\biggl(& \delta ^{m+n-r}_c(f_{ab}^c[\beta
^b,\alpha ^a]^r_+t_c)+\delta ^{m+n-r}(d_{ab}^c[\beta ^b,\alpha
^a]^r_-t_c) \cr +
& 2\widetilde \delta ^{m+n-r}([\beta ^a,\alpha _a]^r_-)
\biggr)g(x),\cr}}
where the following notations are used
\eqn\efda{2\lbrack \beta _n,\alpha _m\rbrack ^r_{\pm }=
\pmatrix{ n \cr r \cr }\beta _n\partial ^r\alpha _n\pm
\pmatrix{ m \cr r \cr }\alpha _m\partial ^r\beta _n}
and it is taken into account that the binomial coefficients
\ $\pmatrix{m \cr r \cr }=0$ \ for \  $r>m$.

We note that the nonabelian transformations \efb \ alone do not
form a closed algebra (see \rAA ). We shall have a closed algebra
if we include the extension of the abelian gauge transformations
($W$-transformations) which we find from \efb \ by replacing
$\hat \alpha $ with $k$, i. e.
\eqn\efaz{ \widetilde \delta ^m(k)g(x)=
ik_m(x)\partial ^{m+1}g(x), \qquad  (m=0,1,\dots).}

Consequently, to have a closed  Lie algebra we must consider
also the commutators:
\eqn\eff{[\widetilde \delta ^m(k),\delta ^n(\hat \beta )]g(x)=
\sum ^{max (m+1,n)}_{r=0}\delta ^{m+n-r+1}([\hat \beta
,k]^r_-)g(x).}
and
\eqn\etr{[\widetilde \delta ^m(k),
\widetilde \delta ^n(h)]g(x)=
\sum ^{\max(m+1,n+1)}_{r=0}\widetilde \delta ^{m+n-r+1}
\Bigl([h_n,k _m]^r_-\Bigr)g(x).}

The commutation relations \efd , \eff \ and \etr \ show that
the transformations \efabr \ and \efb \ satisfy a closed Lie
algebra.

The currents which correspond to the transformations \efb \ and
\efaz \ are given by:
\eqn\efg{J^m_a(x)=tr\{\partial g^{-1}(x)\partial ^mg(x)t_{a}\}
, \quad (m=0,1,\dots ),}
and
\eqn\etmm{V^{m} = -tr\bigl(jU^{m+1}\bigr)=
tr\bigl(\partial g^{-1}\partial ^{m+1}g\bigr), \quad (m=0,1,\dots ),}
which are conserved if the corresponding equations of motion are
satisfyed, i.  e. \  $\partial _-(g^{-1}\partial _{+}g) = 0$. \
We note, that we start from the algebra (for definiteness) \
$SU(N)\otimes U(1)$ \ which after the introduction of the higher
spin extension is no more a direct product.  Consequently, the
Lie algebra defined by
\efd , \eff \ and \etr , which we call $KW_\infty $ algebra
contains the $\widetilde W_\infty $ algebra
as a higher spin subalgebra.
We note that the r.h.s. of \etr \ differ from the ordinary
$W_{\infty }$ algebra  because  it contains  the   terms \
 $\delta ^{m+n}g,\ \delta ^{m+n-1}g, \dots, \delta ^{\min (m,n}$
\ instead of \ $\delta ^{m+n}g, \delta
^{m+n-2}g,\dots , \delta ^{(0,1)}g$  \ for  the case of \ $W_{\infty
}$ \ algebra \rP . The algebra \etr , which we denote by
\ $\widetilde W_\infty $ \ admits a central extension for any spin
which, however, is nondiagonal. As an example the  first
nondiagonal central  charge term is given by \ $C^{1,0}_{mn}=\delta
_{m+n,0}cm^{2}(1-m^{2})/24$ \ where \ $c$ \ is the Virasoro central
charge. However, by  suitable redefinition of the transformations
\etr , and as a consequence a redefinition of the corresponding
currents \etmm , we can find these quantities  in the basis used
in \rP \ for the \ $W_{\infty }$ \ algebra with diagonal central
charges.

The transformations generating the \ $W_\infty $ \ algebra currents in the
ordinary basis used in \rP \ can be found from \efaz \ by a simple
redefinition
\eqn\etg{\widetilde {\delta }^mg(x)=
\sum ^m_{l=0}a_m\partial ^lk_m(x)\partial ^{m-l+1}g(x),}
where \ $a_{m}$ \ are constant parameters (see \rP ). We  note  that
if \ $m \neq 1$ \ the transformations \etg \ form only an on-shell
symmetry  of  the  action \ea , however, these
transformations also generate conserved currents
\eqn\eti {\widetilde V^m=
\sum ^m_{l=0}\tilde a_ltr(\partial ^{l+1}g^{-1}\partial^{m-l+1}g),
\qquad (m=0, 1, 2, \dots ).}
It is easy to show also  that \eti \ and \etmm\ are
connected by a simple redefinition
\eqn\etk {\widetilde V^m(x)\rightarrow V_m(x)+
\sum ^m_{r=1}b^m_r\partial^rV^{m-r}(x),\qquad (m=0,1,2,\dots ),}
where $b$ are suitable constants.\foot{The current j \eaaa
\ is transformed with respect to the linear transformations \etg
\ (if $m\ge 1$) according to a nonabelian gauge law \ebbb \ for
arbitrary choice of the parameters $a_l$, $l\ge 1$.} The same can
be done for the "nonabelian" transformations \efb \ and for the
corresponding currents \efg .

The currents \efg \ and \etmm \ in general (if $m\ge 1$) do not form a
closed Lie algebra with respect to the Poisson bracket deffined
in \rW .

\newsec{Nonlineal realization}

\noindent The characteristic property of the WZWN model is  that
there is an equivalence between  linear and nonlinear
realizations  of the $KW_\infty $  algebras  on  the  classical
level.\foot{The nonlinear realization of $W_\infty $ can be found
from $w_\infty $ by quantum deformation.} To demonstrate this
statement we return to the formula \efaa \ which shows
that  both linear and nolinear spin-$2$ gauge
transnsformations are equivalent. To show this equivalense for
the higher spin transformations we consider the commutator for
two nonlinear spin-$2$ gauge transformations $[\widetilde \delta
^1_a,\widetilde \delta ^1_b]g$. The l. h. s.  of this commutator
coincides (by the form) with the l. h. s. of the formula \efabr \
if we set
\eqn\efkk{\delta ^2(\hat \alpha )g(x)=
i\alpha ^a(x)g(x)(j^2+\partial j)t_a}
and
\eqn\efkl{\widetilde \delta (k)g(x)=ik(x)g(x)j.}
{}For convenience we introduce the notation
\eqn\efkm{\widetilde U^2=(D^2)=j^2+\partial j,}
where
\eqn\efkn{D=j+\partial ,}
and $\partial $ is a right-acting derivative. It is easy to see
that \efkm \ is equivalent to
\eqn\efko{\widetilde U^2=U^2=g^{-1}\partial ^2g,}
where $U^2$ is given by \efbn .

To demonstrate this eqivalence for arbitray spin we use
the following useful recursive relation
\eqn\etl{\widetilde U^{m+1}=(j+\partial )\widetilde U^m=
(D\widetilde U^m),\qquad  (m=0,1,2,\dots ),}
Starting from $\widetilde U^{0}=I$ we obtain
\eqn\erm{\widetilde U^m=(D^m), \qquad \quad (m=1,2,\dots )}

The formulas \etl \ and \erm \ allow us to write down the following
equalities for the linear and nonlinear realizations for arbitrary
spin
\eqn\efla{\widetilde U^m=U^m=g^{-1}\partial ^{m}g.}

The nonlinear realization of $KW_\infty $ and the
$\widetilde W_\infty $ transformations have the following form:
\eqn\eflb{\delta ^m_a(\hat \alpha )g(x)
=i\alpha ^ag(x)\tilde U^mt_a,\qquad (m=0,1,2,\dots )}
\eqn\eflc{\widetilde \delta ^m(k)g(x)
=ik(x)g(x)\tilde U^{m+1}, \qquad (m=0,1,\dots ).}

The transformations \eflb \ and \eflc \ satify the commutation
relations \efd , \eff \ and \etr , i.e. the same algebra as the
corresponding linear transformations.

Taking into ackount \efla \ we can conclude that both the linear
\efb , \efaz \ and the nonlinear \eflb ,\eflc \ transformations are
equivalent. We note that this property is characteristic for the
WZNW model only.  The similar relations between the coresponding
conserved currents
\eqn\efqq{\widetilde J_a^m(x)=
tr(j\widetilde U^{m}t_a), \qquad (m=0,1,2,\dots )}
\eqn\efqr{\widetilde V^m(x)=tr(j\widetilde U^{m+1}),
\qquad (m=0,1,2,\dots )}
also take place.

Now we shall show that the currents \efg \ and \etmm \
(\efqq \ and \efqr ) do not form an invariant space with respect to
the transformations  \efb \ and \efaz (\eflb \ and \eflc ).
Indeed considering for simplicity the
bilinear realization for $V^m$ we have:
\eqn\esh{\eqalign{
\delta ^mV^n(x) & =tr\bigl(\partial (\delta ^mg^{-1})\partial ^{n+1}g +
\partial g^{-1}\partial ^{n+1}(\delta ^{m+1}g)\bigr) \cr
& =-\partial \Bigl(k_mtr\bigl(g^{-1}\partial ^{m+1}gg^{-1}
\partial ^{n+1}g\bigr)\Bigr)+
k_mtr\bigl(g^{-1}\partial ^{m+1}gg^{-1}\partial ^{n+2}g\bigr) \cr
& +\sum_{r=0}^{m+1}\pmatrix{m+1 \cr r \cr }
\partial ^rk_mtr\bigl(\partial g^{-1}\partial ^{m+n-r+2}g\bigr),\cr }}
where we  use the transformation law for $g^{-1}$:
\eqn\esi{\delta ^mg^{-1}=-g^{-1}\delta ^mg g^{-1}=
-k_mg^{-1}\partial ^{m+1}gg^{-1},}
which satisfies the same Lie algebra \etr .
In the case $m=1$ corresponding to the conformal
transformations we find from \esh \
\eqn\eshh{\delta ^0V^n(x)=(n+2)\partial k_0V^n(x)+
k_0\partial V^n+\sum _{r=2}^{n+1}\pmatrix {n+1 \cr r \cr }
\partial ^rk_0V^{n-r+1},}
which shows that only for $n=0$ we have a primary field
transformation law and for $n>0$ we have quasi-primary
transformation law.

We note that in the general case $m>0$ in the formula \esh \
there appear the fourth order (with respect to g) terms due to
which the currents \efg \ and \etmm \ do not form an invariant
space.  It can be shown that in the general case $m>0$ it is
impossible to reduce these higher degree terms into the bilinear
ones only.  To check the latter statement we consider the first
term in the second line of \esh \ for the simplest notrivial case
$m=n=1$
\eqn\esj{tr\{g^{-1}\partial ^2gg^{-1}\partial
^2\}= V^2-\partial V^1-tr\{g^{-1}\partial g\partial
g^{-1}\partial ^2g\}.}

It is impossible to express the last term in \esj \ as a linear
combination of the bilinear currents only. To clarify this
observation we return to the transformation law \eflc \ from
which we find (for $k_m(x)=1$)
\eqn\esja{\eqalign{\delta ^m(\delta ^ng) & =\delta ^m(gU^{n+1})=
\delta ^mgU^{n+1}+g\delta ^mU^{n+1} \cr
& =g(U^{m+1}U^{n+1}-U^{m+1}U^{n+1}+U^{m+n+2})=\delta ^{m+n+1}g,
\cr }}
where
\eqn\esjb{\delta ^mU^{n+1}=U^{m+n+2}-U^{m+1}U^{n+1}.}
The second term in \esjb \ can be represented linearly with
respect to $U$ only if $m=1$. The formula \esja \ shows  that
the nonlinear term which appears in the current transformation law
\esjb \ is canceld in the field transformation law. The latter
gives a possibility to close an algebra in linear realization
also in the case
when the currents space is noninvariant. It is possible to have
an
invariant currents space if the nonlinear transformation appears
also in
the transformation laws (see formula \ef ) which lead in the
general case to some $W_{\infty \times \infty }$ algebra. For the
nonabelian case, which we  consider in the present article, the
inclusion of nonlinear transformations, in general case, leads to
additional explosion of the algebra.

The same property possess the nonlinearly realized currents \efqq and
\efqr .

The noninvariance of the currents spaces \efg \ and \etmm
, as well as, \efqq \ and \efqr \
 makes it imposible to  gauge the symmetries under consideration.

We note, that the above discussed nonivariances of the
currents spaces with respect to their generating transformations is
a characteristic property of the WZNW model, as well as of the
principal chiral model. This is a consequence of the
transformation law for the field $g^{-1}$ \esi . The latter is
the difference between the free-fermionic model (see \rAA ) and
the WZNW model which on the ordinary symmetry level are
completely equivalent \rW .

We note also, that in the
case of the free fermionic nonabelian model \eb \ which is on-shell
invariant with respect to the extended gauge transformations
\eqn\esk{\delta \psi =i\sum_{l\ge 0}k_l\partial ^{l+1}\psi ,}
the corresponding conserved currents
\eqn\esl{J^l=i\sum _{m=0}^N\psi _m\partial ^{l+1}\psi _m }
form invariant space (up to a total derivative term) with respect
to \esk \  which is easy to be checked.

\newsec{Nonlocal realization}

\noindent We recall that in the general case of sigma models
there exist also nonlocal off-shell symmetry transformations
giving  nonlinear realisation of the affine Kac-Moody algebra
(see \rD , \rZb \ and references there). All conserved currents
have spin equal to 1 as a consequence of nonlocality. The
starting point to find these symmetry transformations is the
following equation for the generators
\eqn\efh{\partial T_a(x,\lambda)=
{\lambda \over 1-\lambda}[j(x),T_a(x,\lambda )]}
with the boundary condition
\eqn\efi{T_a(x,\lambda )\vert _{x=-\infty }=t_a,}
where $\lambda $  is dimensionless parameter. For our purpose the
following representation is usefull \rWu
\eqn\efj{T_a(x,\lambda ) =
\sum\limits_{k=-\infty }^\infty \lambda ^kT_a^k(x) =
\chi(x,\lambda )t_a\chi ^{-1}(x,\lambda ).}
It is easy to verify that from equation \efh \ it follows an
equation for $\chi $
\eqn\efk{\partial \chi (x,\lambda )=
{\lambda \over 1-\lambda }j(x)\chi(x,\lambda ),}
with the boundary condition
\eqn\efl{\chi (x,\lambda )\vert _{x=-\infty }=I.}
The solutions of the equation
\efk \ satisfing the boundary condition \efl \ are given by:
\eqn\efm{\chi ^{k+1}(x)
=\chi ^k(x)+\int\limits_{-\infty}^x j(y)\chi ^k(y)dy,}
for $k=0,1,\dots$, and
\eqn\efn{\chi ^{-k}(x)=
-g^{-1}(x)\int\limits_{-\infty}^xdyg(y)\left(\partial
+j(y)\right)\chi ^{-k+1}(y),}
for $k=1,2,\dots $.

As follows from \eac , \efj , \efm \ and \efn \ the nonlocal
gauge transformations\foot{In our considerations $\alpha $ in
generaly depends on $x$. This is the difference from
the papers \rD , \rZa , \rZb \ and \rW \ where
$\alpha $ is a constant.}
\eqn\efo{\delta _{\alpha ,\lambda }g=gT_a(x,\lambda )\alpha^a(x),}
where $\alpha $ depends on one of the light-cone
coordinates only, are at least on-shell symmetry of the action
\ea . Now, we shall show that if $\alpha $  are constant
parameters then we have also off-shell symmetry.  Indeed
incerting \efj \ in \eac \ we find

\eqn\efp{\eqalign{\delta S & =
\int d^2xtr\{j_+(x)\partial _{-}
\Bigl(T_a(x,\lambda)\alpha ^a(x)\Bigr)\}\cr
& ={(1-\lambda )\over \lambda }\int
d^2x\alpha ^atr\Bigl( -\partial _{-}\chi (x,\lambda )t_a\partial
_+\chi ^{-1}(x,\lambda )\cr
&+\partial _+\chi (x,\lambda
)t_a\partial _-\chi ^{-1}(x,\lambda )\Bigr)\cr
& =\epsilon ^{\mu \nu }\int d^2x\alpha ^a(x)tr\Bigl( \partial _\mu
\chi (x,\lambda )t_a\partial _\nu \chi ^{-1}(x,\lambda)\Bigr),\cr }}
where we use the equation \efk \ and the corresponding
equation for $\chi ^{-1}$ which we obtain from equation \efk \ using
 the identity $\chi \chi^{-1}=1$.
Consequently, from \efp \ it follows that the transformations
\efo \ are off-shell symmetry of the action \ea \ only if
$\alpha $ are constant parameters. We recall, that in the latter
case the transformations \efo \ give nonlinear realisation of the
affine Kac-Moody algebra which are also  nonlocal \rD , \rWu , \rZa .
We obtain the Lie algebraic structure of the nonlocal gauge
transformations
\efo \ considering the commutator:
\eqn\efq{\lbrack \delta _{\alpha ,\lambda },\delta _{\beta ,\alpha
}\rbrack g(x)
=g(x)\{ \lbrack T(\alpha ,\lambda ),T(\beta ,\tau )\rbrack
+\delta _{\alpha ,\lambda }T(\beta ,\tau )-\delta _{\beta
,\tau }T(\alpha ,\lambda )\},}
where we denoted $T(\alpha ,\lambda )=T_a(x,\lambda )\alpha
^a$.  We find the variations $\delta T$ as
\eqn\efr{\delta _{\alpha ,\lambda }T(\beta ,\tau )
=\{ \delta _{\alpha ,\lambda }\chi (\beta ,\tau )t_b\chi
^{-1}(\beta ,\tau ) + \chi (\beta ,\tau )t_b\delta _{\alpha
,\lambda }\chi ^{-1}(\beta ,\tau )\}\beta^b.}
$\delta \chi $  is a  solution of the equation :
\eqn\efs{\eqalign{
\partial\{\delta _{\alpha ,\lambda )}\chi (x,\tau )\}
 &={\tau \over (1-\tau )}\{\delta _{\alpha ,\lambda
}j(x)\chi (x,\tau )+j(x)\delta _{\alpha ,\lambda }\chi (x,\tau
)\}\cr
 &={\tau \over (1-\tau )}\{\partial \left(T(\alpha ,\lambda
)+\lbrack j(x),T(\alpha ,\lambda )\rbrack \right)\chi (x,\tau
)\cr
&+j(x)\delta _{\alpha ,\lambda }\chi (x,\tau )\}\cr}}
which we derived from \efk .  The solution of \efs \ satisfying
the boundary condition
\eqn\eft{\delta \chi (x,\lambda )\vert _{-\infty }=0,}
is given by
\eqn\efu{\eqalign{
\delta _{\alpha ,\lambda }\chi (\beta ,\tau )
& ={\tau \over (\lambda -\tau )}\{ T(\alpha ,\lambda )\chi (\tau )
-\chi (\tau )t_a\alpha ^a(-\infty )\}\cr
& -{\tau (1-\lambda )\over (1-\tau )}(\lambda -\tau )\chi (x,\tau )
I(x,\alpha ,\lambda ,\tau ).\cr}}
Here we introduced the notation
\eqn\efv{I(x,\alpha ,\lambda ,\tau )=\int _{-\infty }^xdy\chi
^{-1}(y,\tau )T_a(y,\lambda )\chi (y,\tau )\partial \alpha (y).}

Inserting \efu \ into \efr \ and \efq \ we obtain:
\eqn\efw{\eqalign{
\lbrack \delta _{\alpha ,\lambda },\delta _{\beta ,\tau }\rbrack
g(x)
& = {f_{abc}\over (\lambda -\tau )}\Bigl(\tau \alpha ^a(-\infty)
 \beta ^b(x)\delta _{c,\tau }g(x)
-\lambda \alpha ^a(x)
 \beta ^b(-\infty )\delta _{c,\lambda }g(x)\Bigr)\cr
& -{1\over (\lambda -\tau )}g(x)\{\tau \chi (\tau )
 \lbrack I(x,\alpha ,\lambda ,\tau ),t_b\rbrack \chi ^{-1}(\tau )
 \beta ^b(\tau )\cr
& +\lambda \chi (\lambda )\lbrack I(x,\beta ,\tau ,\lambda ),t_a\rbrack
 \chi ^{-1}(\lambda )\alpha ^a(x)\}.\cr}}
{} From \efw \ it follows that for the zero modes $\alpha =\beta
=const$ we have the ordinary nonlocal and nonlinear realization
of the affine algebra. In all other cases, if \ $\alpha
(-\infty ), \beta (-\infty )$ \ are finite, we
have an open algebra.

\newsec{Conclusions}

\noindent The higher spin extension of the affine Kac-Moody
algebra show us that we have closed Lie algebra only if we start
with \ $U(N)$ \ or \ $GL(N)$ \ spin \ $1$ \ affine Kac-Moody
algebras. In the case when we start with the \ $O(N)$ \ algebra
the extended algebra closes to the extended \ $GL(N)$ \ algebra.
The corresponding higher spin extended algebra contains as a
subalgebra the classical \ $W_\infty $ \ algebra. The obtained
linear and nonlinear realizations of these extended algebras are
equivalent, which is a characteristic property of the WZNW model.
Another property of the WZNW model is that the considered here
higher spin conserved quantities do not form invariant space,
which makes impossible for the corresponding symmetry to be
gauged. This allows us to conclude that there do not exist gauge
theory based on the higher spin extended affine Kac-Moody
symmetry considered here for the WZNW model.  Our considerations
allow us to conclude also that the equivalence between the
nonabelian free fermionic model and the WZNW model is violated on
the higher spin symmetry level.

\noindent{\bf Acnowledgements}

\noindent The author would like to thank to Prof. Abdus Salam for
the hospitality in the ICTP in Trieste where the present
article was initiated and to Prof. E. Sezgin for stimulating
discussion on the early stage of these investigations.

\listrefs